\newcommand{\beq}{\begin{equation}} \newcommand{\eeq}{\end{equation}}
\newcommand{\bea}{\begin{eqnarray}} \newcommand{\eea}{\end{eqnarray}}
\newcommand{\bear}{\begin{eqnarray*}} \newcommand{\eear}{\end{eqnarray*}}
 
\newcommand{\rf}[1]{(\ref{#1})}

\documentclass[a4paper,12pt]{article}
\usepackage{amsmath}
\usepackage{amsfonts}
\usepackage{amssymb}
\usepackage{epstopdf}
\usepackage[dvips]{graphicx}

\begin{document}

\title {The  matrix product ansatz for the six-vertex model}

\author{Matheus J. Lazo \\ {\small Universidade de S\~ao Paulo, Instituto de F\'{\i}sica de S\~ao Carlos,} \\{\small Caixa Postal 369, 13560-590 S\~ao Carlos, S\~ao Paulo, Brazil}}


\maketitle

\begin{abstract}

Recently it was shown that the eigenfunctions for the the asymmetric
exclusion problem and several of its generalizations as well as a huge 
family of quantum chains, like the anisotropic Heisenberg model,  
Fateev-Zamolodchikov model,  Izergin-Korepin model, Sutherland model,  
$t-J$ model, Hubbard model, etc, can be expressed by a matrix 
product {\it ansatz}. Differently from the coordinate Bethe {\it ansatz},
where the eigenvalues and eigenvectors are plane wave combinations, in this
{\it ansatz} the components of the eigenfunctions are 
obtained through the algebraic properties of properly 
defined matrices.
In this work, we introduce a formulation of a matrix 
product {\it ansatz} for the six-vertex model with periodic boundary condition, which is the paradigmatic
example of integrability in two dimensions. Remarkably, our studies of the
six-vertex model are in agreement with the conjecture that all models exactly
solved by the Bethe {\it ansatz} can also be solved by an appropriated matrix
product {\it ansatz}.  

\end{abstract}


\section{Introduction}

The Bethe {\it ansatz} in its several formulations (coordinate, inverse
scattering and functional) has been established over the years as a powerful
tool for the description of the eigenvectors of a huge variety of integrable
one-dimensional quantum spin chains and two-dimensional transfer matrices (see e.g. \cite{baxter}-\cite{revschlo}  
for reviews). A quantum Hamiltonian or a transfer matrix is considered exactly
integrable if an infinite number of its eigenstates can be expressed by the
Bethe {\it ansatz} in the thermodynamic limit.
In the last two decades it has been shown that
a matrix product {\it ansatz} (MPA) can be used to express the stationary
distribution of the probability densities of some special stochastic
models \cite{derr1}-\cite{alcrit1}. Although these models are in general not
integrable through the Bethe {\it ansatz}, they have the components of its
ground-state wavefunctions given in terms of a product of matrices. According
to this {\it ansatz}, the algebraic properties of the matrices defining the
MPA fix the components of the wavefunction apart from
a normalization constant. 

An important development of the MPA that appeared in
the context of stochastic models is the dynamical matrix product {\it ansatz} (DMPA) 
\cite{stinchshutz,reviewshutz}. This DMPA was shown originally to be valid to
the problem of asymmetric diffusion of particles on the lattice
\cite{stinchshutz} and extended to other stochastic models and related spin
Hamiltonians \cite{sasamowada1,popkov}. This {\it ansatz } allows the calculation of the
probability densities, of the stochastic system, at arbitrary times. In the
related spin Hamiltonian this DMPA asserts that not only the ground-state wave
function, as in the standard MPA, but an arbitrary
wavefunction have its components expressed in terms of a matrix product {\it
ansatz} whose matrices, in distinction of the standard MPA, are now time dependent. 

More recently \cite{alclazo1}-\cite{alclazo3}
it was shown that several exactly solvable Hamiltonians, related or not to
stochastic models, may also be solvable by an appropriate time independent
matrix product {\it ansatz}. In this new MPA not only the ground-state
but all wavefunctions can be expressed by a product of matrices. Using this new MPA it was
possible to rederivethe results previously obtained through the Bethe {\it ansatz}
for several quantum chains with one and two global conservation laws, such as
the XXZ chain, spin-$1$ Fateev-Zamolodchikov model, Izergin-Korepin model,
Sutherland model, $t$-$J$ model, Hubbard model, etc \cite{alclazo1,alclazo2}, as well as the exact
solution of the asymmetric exclusion problem with particles of arbitrary size
\cite{alclazo3}. Moreover, the components of the eigenfunctions of the exact
integrable Hamiltonians, which according to the Bethe {\it ansatz} are  normally
given by a combination of plane waves, can also be obtained from the
algebraic properties of the matrices defining the new MPA. In the case of Bethe {\it
ansatz} solutions the eigenvalues and the amplitudes of the plane waves are fixed apart from a
normalization constant by the eigenvalue equation of the Hamiltonian. On the
other hand, in the new MPA, the eigenvalue equation fix the commutation
relations of the matrices defining the {\it ansatz}. The advantage of the new
MPA in the search for new exact integrable models, as showed in our previous works
\cite{alclazo1}-\cite{alclazo3}, is its simplicity and unifying
character in the implementation for arbitrary systems.
All the previous successful applications of the new MPA 
\cite{alclazo1}-\cite{alclazo3} was concerned with the eigenspectrum of
quantum Hamiltonians and stochastic models. In this paper we are going to show
that these results can also be extended to transfer matrix calculations of
two-dimensional classical spin models. More specifically we are going to
extend the new MPA introduced in \cite{alclazo1}-\cite{alclazo3} for the case of
the row-to-row transfer matrix of the six-vertex model with toroidal boundary
condition, which is the paradigm of integrability in two dimensions. This
transfer matrix was diagonalized through the coordinate Bethe {\it ansatz}
firstly by Lieb \cite{lieb}, in a special case, and by Sutherland
\cite{sutherland} and Yang \cite{Yang,Yang2}.


\section{The asymmetric six-vertex model and its transfer matrix}

The six-vertex model defined on a square lattice, was introduced to explain the residual entropy of the ice
\cite{lieb}-\cite{Yang2}. We are going to consider the asymmetric version of
the six-vertex model that was first studied and exactly solved with standard
methods \cite{Nolden,BukmanShore}. This model is defined on a square lattice with $M$
rows and $N$ columns and toroidal boundary condition. At each horizontal (vertical) lattice
bond we attach an arrow pointing to the left or right (up or down)
directions. These arrows configurations can be equivalently described by the
vertex configurations of the lattice. A vertex configuration at a given site
(center) is formed by the four arrows attached to its links. The allowed
vertex configurations are those satisfying the ice rules: two of the arrows
pointing inward and the other two pointing outward of its
center. There are six possible configurations for the vertices. Theses
configurations are showed in fig. 1a. In fig. 1b, a more convenient notation
is introduced, in which we only draw by a solid line (broken line) the links having arrows
pointing to the left or down (right or up) of the center defining the
vertex. Labeling the $M$ rows sequentially by $m=1,2,...,M$ and by $\{x^m \}$
the solid line positions on the vertical edges of the row $m$,
the partition function can be written as 
\beq
\label{0}
Z=\sum_{ \{ x^1 \}}\sum_{\{ x^2 \} }\cdots \sum_{\{ x^M \}} T(\{ x^1 \},\{ x^2
\}) T(\{ x^2 \},\{ x^3 \} )\cdots T(\{x^M \},\{x^1 \} ) = \mbox{Tr}( T^M ),
\eeq
where $T$ is the $2^N \times 2^N$ transfer matrix, with elements
\beq 
\label{1} 
T(\{y\},\{x\}) = \sum e^{-\beta(n_1\epsilon_1+n_2\epsilon_2+\cdots +n_6\epsilon_6)},
\eeq
where the summation is over all allowed arrangements of lines on the
horizontal edges and $n_j$ ($j=1,...,6$) are the numbers
of vertices of types ($1,...,6$) formed by the configurations. For convenience we label the 
Boltzmann weights $a_0,a_1,b_1,b_2,c_1,c_2$ associated with the vertices as in fig. 1. 
It is also important to mention that the number of vertical and horizontal
lines is conserved, forming continuous non-crossing paths through
the lattice. On the other hand the transfer matrix, due to the toroidal
boundary condition, is translation invariant. As a
consequence of these symmetries the transfer matrix breaks up into 
blocks of disjoint sectors labeled by the number $n$ of vertical lines
($n=0,...,N$) and the momentum eigenvalues $P$ ($P=\frac{2\pi}{N}l,l=0,1,...,N-1$).


\section{The Matrix Product {\it ansatz} for the six-vertex model}

The {\it ansatz} we propose \cite{alclazo1}-\cite{alclazo3} states that any eigenfunction $|\Psi_{n,P}\rangle$ of the
transfer matrix \rf{1} in the sector with $n$ ($n=0,1,2,\ldots,N$) vertical lines and momentum
$P$ ($P=\frac{2\pi}{N}l,l=0,1,...,N-1$) is given in terms of a matrix product, i. e., their 
amplitudes are given by the trace of the following  matrix product: 
\beq 
\label{2}
|\psi_{n,P}\rangle = \sum_{x_1,\ldots,x_n}^* \mbox{Tr} (E^{x_1-1}AE^{x_2-x_1-1}A
\cdots E^{x_n-x_{n-1}-1} A E^{L-x_n } \Omega_P) |x_1,\ldots,x_n\rangle,
\eeq
where $|x_1,\ldots,x_n\rangle$ denote the configurations with vertical lines at positions ($x_1,\ldots,x_n$)
and the symbol ($*$) in the sum means the 
restriction to the configurations where $L \ge x_{i+1}> x_i \ge 1$. The objects $A$,
$E$ and $\Omega_P$ are abstract matrices, or operators, with an associative
product whose commutation relations will be fixed by imposing the validity of the eigenvalue equation of the transfer
matrix \rf{1}. The matrices $A$ and
$E$ are associated with the sites where we have a vertical line or not,
respectively, and the matrix $\Omega_P$ is introduced in order to fix the
momentum $P$ of the eigenfunction $|\Psi_{n,P}\rangle$. The fact that $|\psi_{n,P}\rangle$ has a momentum $P$ imply that the 
ratio of the amplitudes corresponding to the configurations $|x_1,\ldots,x_n\rangle$ and 
$|x_1+1,\ldots,x_n+1\rangle$ is $e^{-iP}$, i.e.,
\beq
\label{4}
\frac{\mbox{Tr} (E^{x_1-1}AE^{x_2-x_1-1}A
\cdots E^{x_n-x_{n-1}-1} A E^{L-x_n } \Omega_P)}{\mbox{Tr} (E^{x_1}AE^{x_2-x_1-1}A
\cdots E^{x_n-x_{n-1}-1} A E^{L-x_n -1} \Omega_P)}=e^{-iP},
\eeq
and consequently from \rf{2} we
obtain the following commutation relations 
\beq \label{3}
A\Omega_P = 
e^{-iP}\Omega_PA, \;   E\Omega_P = e^{-iP}\Omega_PE. 
\eeq

The matrix product {\it ansatz} will be valid if the
algebraic relations among the matrices $A$, $E$ and $\Omega_P$ are consistent
with the constrains imposed by the eigenvalue equation
\beq 
\label{5}
T|\psi_{n,P}\rangle = \Lambda_n|\psi_{n,P}\rangle.
\eeq
To solve this last equation, it is helpful to begin, as usual, by considering
the simple cases where $n=0,1$ and $2$ before considering the general case.

{\it \bf The case n = 0.}

In this case the solution of the eigenvalue equation \rf{5} is trivial
since we do not have vertical lines between two successive rows. There are
only two possible horizontal arrangements either all
bonds have a line or all of them are empty. In this case the vertices are all of type $1$ or type
$3$ (see figure 1) and consequently the eigenvalue is given by 
\beq
\label{6}
\Lambda_0=a_0^N+b_1^N
\eeq
where $a_0$ and $b_1$ are the Boltzmann weights of the vertices of types $1$
and $3$, respectively (see figure 1).

{\it {\bf The case n = 1.}}

We have in this case just one vertical line between two rows. The transfer matrix links a vertical line at position $y$ ($y=1,..,N$)
above a  row to a vertical line at any position $x$ ($x=1,...,N$) under
this row. The elements of the transfer matrix $T(y,x)$ in this sector with
momentum $P$ are given by \rf{1}. They are the product of the Boltzmann weights of the vertex appearing on the row. If the position of the
line $x$ is less (greater) than $y$, the vertex configuration at these sites will be of types
$5$ and $6$ ($6$ and $5$) and all the others vertices will be of types
$3$ ($1$) and $1$ ($3$) depending on whether the vertices are between the positions
$x$ and $y$, or not, respectively. In
the case where $x=y$ these vertices should be of type $4$ or $2$ with all the
remains vertices of type $1$ or $3$ respectively. Consequently the eigenvalue
equation \rf{5} for the transfer matrix \rf{1} associated with the components of
$|\psi_{n,P}\rangle$ \rf{2} with $n=1$ and momentum $P$ give us the relations
\bea 
\label{7}
\Lambda_1 \mbox{Tr} (E^{x-1}AE^{N-x}\Omega_P) =&& \sum_{y=x+1}^{N}a_0^{N-y+x-1}b_1^{y-x-1}c_1c_2{\mbox Tr} ( E^{y-1}A
E^{N-y}\Omega_P) \nonumber +\\
&&\sum_{y=1}^{x-1}a_0^{x-y-1}b_1^{N-x+y-1}c_1c_2{\mbox Tr} ( E^{y-1}A E^{N-y}\Omega_P) \nonumber +\\
&& (a_0^{N-1}b_2+b_1^{N-1}a_1)\mbox{Tr} (E^{x-1}AE^{N-x}\Omega_P) .
\eea
Equation \rf{7} can be simplified in order to express all the matrix products
in terms of a single one. This is possible by exploring the cyclic property of
the trace as well as the commutation relations \rf{3}. This allow us to
factorize the matrix product in the following form
\bea  
\label{8}
\Lambda_1 =&& \sum_{y=x+1}^{N}a_0^{N-y+x-1}b_1^{y-x-1}c_1c_2 e^{-iP(y-x)}+
\sum_{y=1}^{x-1}a_0^{x-y-1}b_1^{N-x+y-1}c_1c_2 e^{-iP(y-x)} \nonumber +\\
&& (a_0^{N-1}b_2+b_1^{N-1}a_1).
\eea
By evaluating the sums in \rf{8} we obtain
\beq 
\label{9}
\Lambda_1 = a_0^N L(P) + b_1^N M(P) + b_1^N \frac{c_1c_2}{a_0}\left( \frac{b_1}{a_0}
\right)^{-x} \frac{e^{-iP(1-x)}}{a_0-b_1e^{-iP}}(1 - e^{-iNP}),
\eeq
where
\beq 
\label{10}
L(P) = \frac{a_0b_2+(c_1c_2-b_1b_2)e^{-iP}}{a_0^2-a_0b_1e^{-iP}} \;\;\;\;\;
\mbox{and} \;\;\;\;\; M(P) = \frac{a_0a_1-c_1c_2-a_1b_1e^{-iP}}{a_0b_1-b_1^2e^{-iP}}.
\eeq 
In order to satisfy \rf{5}, the eigenvalue $\Lambda_1$ in \rf{9} should be
independent of the vertical line position $x$. Thus the last term in the right
hand side of \rf{9} must vanish. The only way to cancel this term, for non
zero Boltzmann weights, is obtained by imposing the following constraint to
the momentum $P$
\beq
\label{11} 
e^{iNP}=1,
\eeq 
which is automatically satisfied, since $P=\frac{2\pi}{N}l$ $l=0,1,...,N-1$. The
eigenvalue \rf{9} is then given by
\beq 
\label{12}
\Lambda_1 = a_0^N L(P) + b_1^N M(P).
\eeq

An alternative solution of \rf{7}, whose generalization will be convenient for arbitrary values
of $n$, is obtained by expressing the matrix $A$ in terms of the matrix $E$
and a spectral parameter dependent matrix
\beq
\label{13}
A=A_kE,
\eeq
with $A_k$ satisfying
\beq
\label{14}
EA_k=e^{ik}A_kE.
\eeq
As a consequence of \rf{3} and \rf{14} $A_k$ also satisfies
\beq
\label{14b}
A_k\Omega_P=\Omega_PA_k.
\eeq
The spectral parameter $k$ will be fixed by the eigenvalue equation \rf{7}.
Inserting \rf{13} in \rf{7} and using the commutation relation \rf{14} we
obtain \rf{8} with the value $k$ replacing $P$. Therefore
\beq 
\label{15}
\Lambda_1 = a_0^N L(k) + b_1^N M(k), 
\eeq
with
\beq
\label{16}
e^{iNk}=1, \;\;\;\;\; k=\frac{2\pi}{N}l \;\;\; (l=0,1,...,N-1).
\eeq
Comparing \rf{11} and \rf{12} with \rf{15} and \rf{16} we observe the equality 
$k=P$. This fact can also be seen directly by inserting \rf{13} in \rf{4}
and using \rf{14}.

We still need to verify whether the algebraic relations among the matrices $A_k$,
$E$ and $\Omega_P$ \rf{3}, \rf{14} and \rf{14b} are consistent with the cyclic
property of the trace. Indeed these equations yield
\bea
\label{17}
\mbox{Tr} (A_kE^N\Omega_P)&&=e^{-iNk}\mbox{Tr}
(E^NA_k\Omega_P)=e^{-iNk}\mbox{Tr} (E^N\Omega_PA_k) \nonumber \\  
&&= e^{-iNk}\mbox{Tr} (A_kE^N\Omega_P),
\eea
which satisfies the cyclicity of the trace due to \rf{16}. Since no new
constraints is obtained for the matrices $A_k$,
$E$ and $\Omega_P$, with $k=P$,  and for spectral parameter $k$, the MPA is consistent.

{\it {\bf The case n = 2.}}

In this sector there are two vertical lines in the row. We have in general two
types of relations, which are relations where at least one of the vertical lines
($y_1$,$y_2$) coincide with ($x_1$,$x_2$) and those where $y_1$ and $y_2$
interlace with $x_1$ and $x_2$ ($x_1<y_1<x_2<y_2$
or $y_1<x_1<y_2<x_2$). Then, the eigenvalue equation \rf{5} imply
\bea 
\label{18}
\Lambda_2 &&\mbox{Tr} (E^{x_1-1}AE^{x_2-x_1-1}AE^{N-x_2}\Omega_P) = \nonumber \\
&& \sum_{y_1=x_1}^{x_2}\sum_{y_2=x_2}^{\;\;\;\;N\;\;*} a_0^{N-y_2+x_1-1}c_2f(x_1,y_1)g(y_1,x_2)f(x_2,y_2){\mbox Tr} ( E^{y_1-1}AE^{y_2-y_1-1}A
E^{N-y_2}\Omega_P) \nonumber +\\
&& \sum_{y_1=1}^{x_1}\sum_{y_2=x_1}^{\;\;\;\;x_2\;\;*} b_1^{N-x_2+y_1-1}c_1g(y_1,x_1)f(x_1,y_2)g(y_2,x_2){\mbox Tr} ( E^{y_1-1}AE^{y_2-y_1-1}A
E^{N-y_2}\Omega_P),
\eea
where the symbol $*$ in the sums means that terms with $y_1=y_2$ are excluded
and 
\beq
\label{19}
f(x,y)=\left\{ \begin{array}{cc}
              \frac{b_2}{c_2} & \mbox{if}\;x=y \\
              c_1b_1^{y-x-1}  & \mbox{if}\;y>x

              \end{array}
       \right.
\;\;\;\;\; \mbox{and} \;\;\;\;\;
g(y,x)=\left\{ \begin{array}{cc}
              \frac{a_1}{c_1} & \mbox{if}\;x=y \\
              c_2a_0^{x-y-1}  & \mbox{if}\;x>y

              \end{array}
       \right..
\eeq               
The relation \rf{18} connects configurations where the arrangements of vertical
lines above one row do not have the same distance of the vertical
lines below this same row. In other words, the distance of the incoming lines $y_2-y_1$ are in
general different of the outcoming distance $x_2-x_1$. As a consequence, it is not possible to solve the
eigenvalue equation by just using the cyclic property of the trace in \rf{2}
as done previously in the case $n=1$. We need now to use a generalization of the
algebraic relation \rf{13} for the case of two lines. The generalization of
\rf{13} is done by writing the matrix $A$ in terms of two new spectral
parameter matrices as:
\beq   
\label{20}                         
A=\sum_{j=1}^2 A_{k_j}E,
\eeq
with the commutation relations 
\beq
\label{21}
EA_{k_j}=e^{ik_j}A_{k_j}E \;\;\;\;\; {\mbox and} \;\;\;\;\;
A_{k_j}\Omega_P=\Omega_PA_{k_j} \;\;\;\;\; (j=1,2),
\eeq
where the spectral parameters $k_1$ and $k_2$ are up to now unknown complex numbers.

Inserting \rf{20} in \rf{18} and using in this expression \rf{21} and \rf{10}
we obtain, after similar manipulation as we did in the case $n=1$, the following constraints
\bea
\label{22}
&&\sum_{j,l=1}^2\left[ \Lambda_2 -a_0^NL(k_j)L(k_l)-b_1^NM(k_j)M(k_l) \right] e^{-ik_jx_1}e^{-ik_lx_2}\mbox{Tr} (A_{k_j}A_{k_l}E^N\Omega_P)\nonumber \\
&&-\sum_{j,l=1}^2 a_0^N\left[ L(k_l)M(k_j)-\frac{a_1b_2}{a_0b_1} \right] \left( \frac{b_1}{a_0}
\right)^{x_2-x_1}e^{-i(k_j+k_l)x_2}\mbox{Tr} (A_{k_j}A_{k_l}E^N\Omega_P)
\nonumber \\
&&-\sum_{j,l=1}^2 b_1^N\left[ L(k_l)M(k_j)-\frac{a_1b_2}{a_0b_1} \right]\left(
  \frac{b_1}{a_0} \right)^{x_1-x_2}e^{-i(k_j+k_l)x_1}\mbox{Tr}
(A_{k_j}A_{k_l}E^N\Omega_P) \nonumber \\
&&+\sum_{j,l=1}^2 b_1^N\frac{c_1^2c_2^2\left[e^{-iNk_l}e^{-ik_jx_1}-e^{-ik_lx_1} \right] e^{-i(k_j+k_l)}}{a_0^2(a_0-b_1e^{-ik_j})(a_0-b_1e^{-ik_l})}\left(
  \frac{b_1}{a_0} \right)^{-x_2}\mbox{Tr} (A_{k_j}A_{k_l}E^N\Omega_P) \nonumber
\\
&&-\sum_{j,l=1}^2 b_1^N\frac{c_1^2c_2^2\left[e^{-i(N+1)k_l}e^{-ik_jx_2}-e^{-ikj}e^{-ik_lx_2} \right] }{a_0b_1(a_0-b_1e^{-ik_j})(a_0-b_1e^{-ik_l})}\left(
  \frac{b_1}{a_0} \right)^{-x_1}\mbox{Tr} (A_{k_j}A_{k_l}E^N\Omega_P)
\nonumber \\
&&+\sum_{j,l=1}^2 b_1^Nc_1c_2b_2\left[\frac{e^{-i(N+1)k_l}e^{-ik_jx_1}
  }{a_0^2(a_0-b_1e^{-ik_l})}- \frac{-e^{-ik_j}e^{-ik_lx_1}}{a_0^2(a_0-b_1e^{-ik_j})}\right] \left(
  \frac{b_1}{a_0} \right)^{-x_2}\mbox{Tr} (A_{k_j}A_{k_l}E^N\Omega_P)
\\
&&+\sum_{j,l=1}^2 b_1^Nc_1c_2a_1\left[\frac{e^{-i(N+1)k_l}e^{-ik_jx_2}
  }{a_0b_1(a_0-b_1e^{-ik_l})}- \frac{-e^{-ik_j}e^{-ik_lx_2}}{a_0b_1(a_0-b_1e^{-ik_j})}\right] \left(
  \frac{b_1}{a_0} \right)^{-x_1}\mbox{Tr} (A_{k_j}A_{k_l}E^N\Omega_P)=0,\nonumber
\eea
where $1\leq x_1 < x_2 \leq N$. This can
only be satisfied if each sum is identically zero. Moreover since $\Lambda_2$
should be independent of $x_1$ or $x_2$ a possible solution of  \rf{22} is
obtained by imposing
\beq
\label{23}
\Lambda_2=a_0^NL(k_1)L(k_2)+b_1^NM(k_1)M(k_2).
\eeq
The algebraic relation between the matrices $A_{k_1}$ and $A_{k_2}$ are obtained
by imposing that both the second and third terms in \rf{22} are zero independently,
i. e., 
\beq
\label{24}
A_{k_j}A_{k_l}=-S(k_j,k_l)A_{k_l}A_{k_j} \;\;\;\;\;(l\neq j)\;\;\;\;\; \left(
  A_{k_j} \right)^2=0 \;\;\;\;\; (j,l=1,2),
\eeq
where
\beq
\label{25}
S(k_j,k_l)=\frac{L(k_j)M(k_l)-\frac{a_1b_2}{a_0b_1}}{L(k_l)M(k_j)-\frac{a_1b_2}{a_0b_1}},
\eeq
with $L(k)$ and $M(k)$ given by \rf{10}. Finally, the vanishing of the last
four terms in \rf{22} will give us relations that fix the spectral parameters
values $k_1$ and $k_2$. These equations are obtained by exploring the algebraic
relations \rf{24} 
\beq
\label{26}
e^{iNk_l}=-S(k_j,k_l) \;\;\;\;\; (l,j=1,2 \;\; \mbox{and} \;\; l\neq j).
\eeq

The eigenvalues and eigenvectors are obtained by inserting the solutions
($k_1,k_2$) of these last equations in \rf{23} and \rf{25}, respectively. The momentum
$P$ is obtained by using \rf{20} and \rf{21} in \rf{4}, i.e., $P=k_1+k_2$.

The consistency of the algebraic equations \rf{20}, \rf{21} and \rf{25} with
the cyclic property of the trace in \rf{2}, as in the case $n=1$, can be
easily verified, yielding
\bea
\label{27}
\mbox{Tr} (A_{k_j}A_{k_l}E^N\Omega_P)&&=-S(k_j,k_l)\mbox{Tr}
(A_{k_l}A_{k_j}E^N\Omega_P) \nonumber \\
&&=-S(k_j,k_l)e^{-iNk_j}\mbox{Tr}(A_{k_l}E^NA_{k_j}\Omega_P) \nonumber \\
&&=-S(k_j,k_l)e^{-iNk_j}\mbox{Tr} (A_{k_j}A_{k_l}E^N\Omega_P).
\eea

{\it {\bf  The case of general n.}}

The previous calculation can be extended for arbitrary values of the number
$n$ of vertical lines. The transfer matrix \rf{1} when applied to the
amplitudes of $|\psi_{n,P}\rangle$ give us an eigenvalue equation linking an
arrangement of vertical lines $x_1,...,x_n$ with arrangements $y_1,...,y_n$ with $x_1\le
y_1\le x_2\le \cdots x_n\le y_n$ and $y_1\le x_1\le y_2\le \cdots y_n\le
x_n$. To solve this eigenvalue equation we need to extend the definition
\rf{20} and the commutation relations \rf{21} for general $n$, i. e.,
\beq
\label{28}
A=\sum_{j=1}^n A_{k_j}E
\eeq
with
\beq
\label{29}
EA_{k_j}=e^{ik_j}A_{k_j}E \;\;\;\;\; \mbox{and} \;\;\;\;\;
A_{k_j}\Omega_P=\Omega_PA_{k_j} \;\;\;\;\; (j=1,...,n),
\eeq
where $k_j $ ($j =1,\ldots,n$) are in general unknown complex numbers that
will be fixed by the eigenvalue equation \rf{5}. Actually the definition
\rf{28} is not the only possible one. The most general definition that enable
us to solve the eigenvalue equation is $A=\sum_{j=1}^n
E^{\alpha}A_{k_j}E^{1+\beta}$, where $\alpha$ and $\beta$ are integer numbers. However
\rf{28} is more convenient since otherwise the S-matrix in \rf{30} and the Bethe
equation \rf{31} will depend on the parameters $\alpha$ and
$\beta$. Inserting \rf{28} in the eigenvalue equation \rf{5} and using the
commutation relations \rf{29} we obtain, similarly as done in the case $n=2$,
the algebraic relations among the matrices
$\{A_{k_j} \}$
\beq
\label{30}
A_{k_j}A_{k_l}=-S(k_j,k_l)A_{k_l}A_{k_j} \;\;\;\;\; \left(
  A_{k_j} \right)^2=0 \;\;\;\;\; (j\neq l=1,...,n),
\eeq
where $S(k_j,k_l)$ is given by \rf{25} and the spectral parameters $k_j$ ($j=1,...,n$) are fixed
by the equation
\beq
\label{31}
e^{iNk_l}=(-1)^{n-1}\prod_{l=1\;\;(l\neq j)}^n S(k_j,k_l) \;\;\;\;\; (j=1,...,n).
\eeq
No new algebraic relations appear for the matrices $\{A_{k_j}\}$ and the 
associativity of the algebra \rf{29} and \rf{30} follows from the property 
$S(k_j,k_l)S(k_l,k_j)=1$. The eigenvalues for the transfer matrix
\rf{1} in the sector with general $n$ is then given by
\beq
\label{32}
\Lambda_2=a_0^NL(k_1)L(k_2)\cdots L(k_n)+b_1^NM(k_1)M(k_2)\cdots M(k_n),
\eeq 
where $L(k)$ and $M(k)$ are given by \rf{10} and the spectral parameters $\{k_j \}$
are the solutions of \rf{31}. The eigenvalues \rf{32} and the spectral
parameter equations coincide with the corresponding equations obtained through
the Bethe {\it ansatz} \cite{Yang,Yang2}.

Finally, the momentum $P$ follows from \rf{4} and \rf{29}:
\beq
\label{33}
P=\sum_{j=1}^n k_j.
\eeq

The consistency of the algebraic relations of the matrices defining the MPA
\rf{2} with the cyclic property of the trace in \rf{2} is promptly verified as
in the cases where $n=1$ and $n=2$. Therefore the MPA is consistent and a
infinite number of eigenvectors of the transfer matrix \rf{1} can be written
by \rf{2} in the thermodynamic limit.


\section{Conclusion}

 In conclusion, we have shown that the new MPA introduced in
 \cite{alclazo1,alclazo2} for one dimensional quantum
 spin chains, such as the XXZ chain, spin-$1$ Fateev-Zamolodchikov model, Izergin-Korepin model,
Sutherland model, $t$-$J$ model, Hubbard model, as well as the exact
solution of the asymmetric exclusion problem with particles of arbitrary size
\cite{alclazo3}, can also be extended to the diagonalization of the row-to-row
 transfer matrix of the six-vertex model with toroidal boundary
 condition. Differently from the standart MPA \cite{derr1}-\cite{alcrit1} this
 new MPA \cite{alclazo1,alclazo2} asserts that all wavefunctions can be
 expressed by a product of matrices. The solution of the six vertex model
 through the new MPA is in  agreement with the conjecture proposed in
 \cite{stinchshutz} and \cite{alclazo1,alclazo2} that all models exactly solved 
 by Bethe {\it ansatz} can also be solved by an appropriate MPA. An
 interesting problem for the future is the formulation of a MPA for others spin
 models like the  $8$-vertex model, which is related to quantum spin chains with no global
 conservation laws such as the XYZ chain.

{\center{acknowledgements: I am grateful to F. C. Alcaraz for his comments and
  M. S. Sarandy for reading the manuscript. This work has been supported by
  CAPES and FAPESP (Brazilian agencies).}}

\newpage

\begin{figure}
\begin{center}
\includegraphics[width=1.0\textwidth]{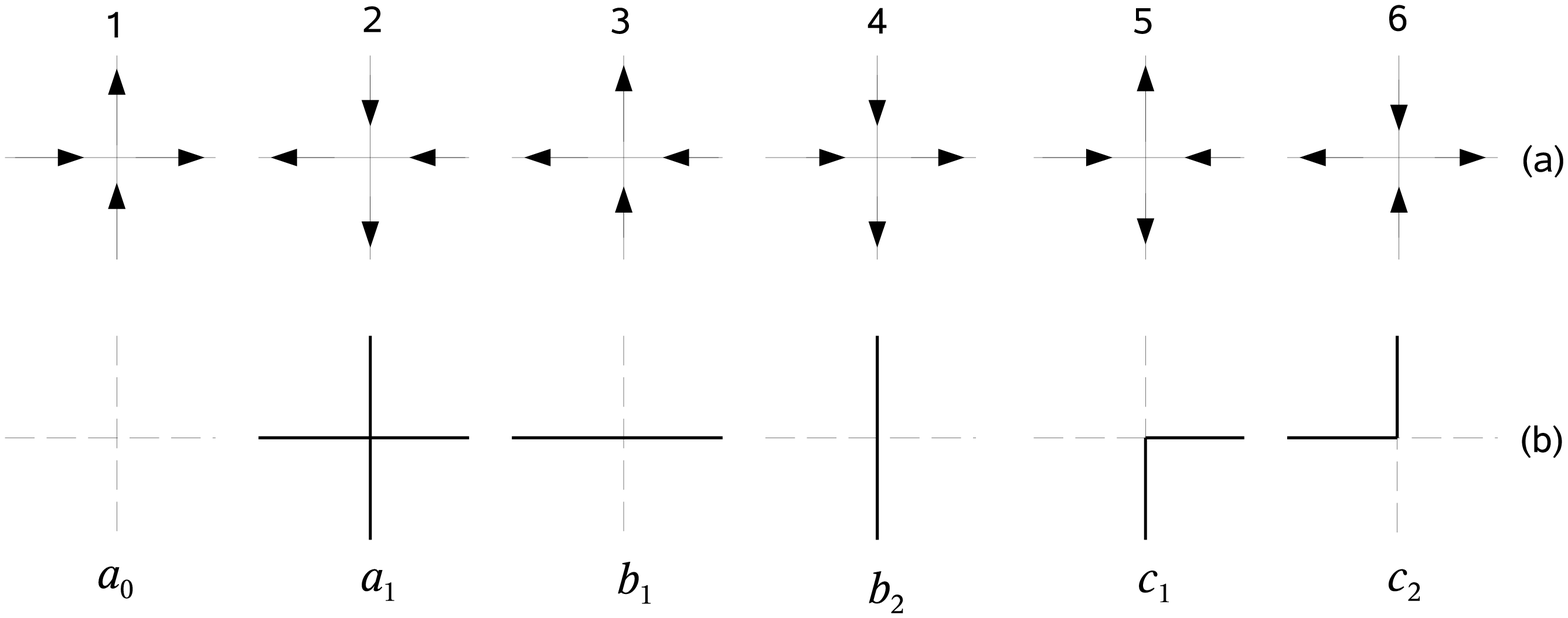}
\end{center}
\caption{The six vertex configurations and their related Boltzmann weights. In
(a) we draw all the arrows and in (b) we draw by solid lines th links where
the arrows are pointing to the down and left directions.}
\end{figure}

\end{document}